\documentstyle[12pt]{article}
\begin{document}
\begin{titlepage}

\begin{flushright}
{\bf MC-TH-95/04}\\
{May 1995}
\end{flushright}

\vspace{1cm}

\begin{center}
\begin{large}
{\bf On $Z$ and $Z + jet$ Production in Heavy Ion Collisions}

\vspace{1cm}

{V.Kartvelishvili$~{}^{a)}$}\\
\bigskip
\end{large}
{\em Department of Physics and Astronomy,\\
Schuster Laboratory, University of Manchester,\\
Manchester M13 9PL, U.K.\\}

\vspace{0.5cm}

\begin{large}
{R.Kvatadze$~{}^{a)}$ and R.Shanidze$~{}^{a)}$}\\
\bigskip
\end{large}
{\em Joint Institute of Nuclear Research,\\
Dubna, Moscow Region, RU-141980, Russia}\\

\vspace{2cm}

{\bf Abstract}
\end{center}

\vspace{0.5cm}

\begin{narrower}
\noindent
$Z$+jet production in heavy ion collisions at the
LHC is proposed as a possible probe of the properties of the
dense hadronic matter. It is shown that the accuracy of this
measurement with general
purpose LHC detectors and under realistic experimental conditions
can be high enough.
It is argued also that $Z$-boson production and subsequent
leptonic decay is a good
reference process for this measurement, as well as for the QCD process
of high $P_T$ jet pair production.
\end{narrower}

\vspace{4cm}

\noindent{---------------------}\\
${}^{a)}$ On leave from High Energy Physics Institute, Tbilisi State
University, Tbilisi, GE-380086, Republic of Georgia.

\end{titlepage}

\newpage
\section{Introduction}
\vspace{0.5cm}

The main physics motivation of heavy ion experiments at LHC is
 possible observation and study of the phase transition from
 the confined hadronic matter to the plasma of deconfined
 quarks and gluons (quark-gluon plasma, QGP).
 At the center of mass energies of few
 TeV per nucleon pair and for heavy nuclei $A \sim 200$,
 the energy density in central nucleus-nucleus collisions must
 be well above the expected phase transition value
 (see e.g. reviews {\cite{tn1,singh}}).

Unfortunately, at present we know no single measurement that could
 give an unambiguous
 indication that the phase transition to QGP has taken place, so
 several observable effects should be measured in order to establish the
 existence of QGP and study its properties. A number of potentially
 measurable effects have been suggested and
 widely discussed in the literature
(see e.g. refs. {\cite{singh,satz2}}).
 A list of those taken from ref.
{\cite{tn2}}, obviously not
 exhaustive and in no particular order, is shown below :
\begin{enumerate}
\item {transverse energy distributions;}
\item {charged multiplicities;}
\item {transverse momentum distributions of identified particles;}
\item {strangeness production;}
\item {identical particle interferometry;}
\item {production of vector mesons with low and high mass;}
\item {continuum of lepton pairs;}
\item {continuum of direct photons;}
\item {high transverse momentum jets.}
\end{enumerate}

Detailed analysis of these topics goes far beyond the scope of this paper,
 but we would like to make some comments. Items 1 to 4 will certainly
 be sensitive to the final state effects, but it seems that reliable
 and unambiguous theoretical predictions about QGP manifestations are difficult
 to obtain {\cite{singh}}.
Same is true for low invariant mass lepton pairs and light vector
mesons, as well as for particle interferometry, with additional
 difficulties for the latter caused by very high
precision of momentum measurements required.
 It also seems that the continuums of high invariant mass
 lepton pairs and direct
 photons will be very hard to study, mainly because of large
 irreducible backgrounds and low rates. Anyway, these two processes can be only
considered as reference processes (see Sect. 3) as final state
interaction should not significantly affect leptons and photons.

In fact, some recent investigations imply that direct probes of
 QGP should be hard enough in order to resolve sub-hadronic scales and
 distinguish confined and deconfined media {\cite{satz}}.
 So, one is essentially left  with two topics most
 widely discussed in the context of QGP manifestations:
 heavy quarkonium production and high transverse momentum jets.
 Yet again, the use of charmonium states ($J/\psi, \psi', \dots$)
at LHC heavy ion experiments may be questionable,
 as their production rates even at average transverse momenta
 should be
 mostly dominated by weak decays of bottom hadrons,
which take place far beyond
 the area where QGP can be formed.
 $\Upsilon$ family production at LHC looks more promising {\cite{gavai}},
 especially
 when the peculiarities of the recent Tevatron data {\cite{Tev}}
 are explained theoretically.

As for high transverse momentum jets,
\begin{equation}\label{jj}
p+p,~A+A \to jet + X,
\end{equation}
 their production mechanisms in
 hadron-hadron collisions are well understood in all accessible ranges of
 the initial energy and jet transverse momentum.
 It is believed that their properties should be affected by the
 dense hadronic matter
{\cite{tn3,pan}}, resulting in measurable effects like
 additional energy loss and acoplanarity.
 Detailed investigation of high $P_T$ jet
 production in heavy ion collisions at LHC and its comparison to that in $pp$
 interactions are well under way {\cite{tn2,jj1,tn12,kk}.
Process (\ref{jj}) is especially interesting
as jet quenching patterns are expected to be different
for hadronic gas and QGP {\cite{tn3,pan}}.

We propose another hard process which also allows to investigate the
 effects of nuclear interaction on the propagation of a fast moving
 color charge, and can potentially distinguish the two states of the dense
 hadronic matter. This is the production of a leptonically decaying $Z$
 boson with high transverse momentum, in association with an energetic
 hadronic jet:
\begin{equation}\label{zj}
p+p,~A+A \to Z(\to \mu^+ \mu^-) + jet + X.
\end{equation}
 The detailed studies of the influence of the dense matter on jet development
 can be found in the literature (see e.g. refs. {\cite{singh,tn3,pan}}).
We do not consider them here;
 the purpose of this paper is just to show that such an investigation is
 theoretically interesting and experimentally feasible at the LHC
 heavy ion facility.
 As shown in Sect. 2,
 the cross section of this process is high enough and the background can
 be easily controlled using physically motivated cuts.
 For the purposes of QGP study, process ({\ref{zj}) incorporates most
 of the useful features of ({\ref{jj}), and
 offers some additional possibilities, which are also discussed in Section 2.

Another problem in heavy ion physics is that
 the cross sections here are not easy to
 normalize properly. This means that instead of comparing a distribution
 in nucleus-nucleus collisions with a similar one obtained in
 nucleon-nucleon interactions, one has to specify an appropriate
 reference process, and
 compare the {\it{ratios}} of two processes.
 In Sect. 3 we argue that $Z$ boson inclusive production with a subsequent
 leptonic decay,
\begin{equation}\label{z}
p+p,~A+A \to Z(\to \mu^+ \mu^-) + X,
\end{equation}
is a good
 choice for the reference process, which satisfies all necessary criteria.

We have performed a Monte-Carlo simulation of the processes
 ({\ref{zj}) and ({\ref{z}), assuming a realistic detector model, which
 can detect $Z \to \mu^+ \mu^-$ decays and high $P_T$ jets in the rapidity
 range $|Y| < 2.5$,
 for lead-lead collisions at LHC
 energies. Some results of our simulation and the discussion of various
 experimental aspects are presented in Sect. 4, and conclusions are
 drawn out in Sect. 5.

\vspace{1.0cm}
\section{$Z + jet$ production}
\vspace{0.5cm}

At multi-TeV accelerators, high transverse energy jet production is
 expected to be completely dominated by the parton subprocess of
 (quasi)elastic gluon-gluon large angle scattering. Hence, in heavy
 ion collisions at the LHC one could investigate the propagation of
 {\it{gluon}}-initiated jets in the dense hadronic matter. The cross
 section of this process is relatively high and large statistics can
 be accumulated. QCD-motivated theoretical expectations are well tested
 at available energies and no serious discrepancies have been found.
 So, the process ({\ref{jj}}) is expected to be the major source of
 information about the effects of the dense matter on the
 fast-moving gluon fragmentation process in heavy
 ion collisions at the LHC. Possible changes in the energy loss
 mechanism in QGP as opposed to dense hadronic gas {\cite{tn3,pan}}
make this process even more interesting.

Nevertheless, some criticism should be expressed:

\begin{itemize}

\item {As all participating partons in the initial hard scattering
subprocess are gluons, it may be rather difficult to distinguish
the effects of the initial state nuclear interactions like gluon
shadowing from genuine final state dense matter effects. In order
to do this, one has to know the gluon distribution functions in heavy
nuclei at small fractional momenta, which would require tedious
analyses of, say, charm and beauty production in lepton-nucleus collisions,
and/or high $P_T$ jet production in {\it{proton\/-nucleus}} interactions.}

\item {No information can be extracted about the influence of the dense
hadronic matter on the fragmentation of {\it{quark}}-initiated jets.}

\item {Both jets in the event are equally affected by the dense hadronic
matter. Most probably, the two jets are affected independently,
which means that certain measurable effects like acoplanarity will
be slightly reduced. But some possible coherent effects may escape
detection.}

\end{itemize}

Consider now the process (\ref{zj}) of a high $P_T$ $Z$-boson production
in association with a jet.
In this case,
relevant parton subprocesses are $q(\bar q) + g \to Z + q(\bar q)$
and $ q+ \bar q \to Z + g$.
Explicit calculations show that in the rapidity region $|Y| < 2.5$ and for
both jet and $Z$ transverse momenta exceeding $50~GeV/c$, the relative
contribution of these two parton subprocesses is $\approx~70~\%$
and $\approx~30~\%$, correspondingly. At least one of the initial partons is
(almost certainly) a quark, and as far as quark distribution functions
are readily measured in high energy lepton-nucleus collisions, initial
state nuclear effects can be taken into account much easier and far more
reliably than for the process ({\ref{jj}}). Moreover, in about 70\% of
events the detected jet takes its origin from a quark or an antiquark,
and by choosing large $|Y|$ values of the $Z+jet$ system this proportion
can be made even higher. This means that the process ({\ref{zj}}) gives
a unique possibility to study the effects of the nuclear matter on
a high purity sample of {\it{light-quark}}\/-induced jets
{\footnote{One could also speculate about
tagging one of the $b(\bar b)$-quark induced jets in the processes
$p+p,~A+A \to b + \bar b + X$ and measuring changes in the
$\bar b (b)$-quark
fragmentation process due to nuclear effects, but experimental
feasibility of this study requires further investigation {\cite{kk2}}.}.

The process (\ref{zj}) has yet another advantage:
the transverse momentum of the jet
is highly correlated to that of the $Z$ (in fact, neglecting the primordial
$P_T$ distribution of initial partons, in the absence of the
initial-state radiation both $P_T$'s are exactly equal and opposite),
so the jet
in this case can be considered as tagged, and the effects of its
further evolution should be easier to determine, as opposed to the two
jet case, where {\it {both}} jets are affected by the final state effects.

{}From the experimental point of view, two high transverse momentum
muons in (\ref{zj}) can be triggered with high efficiency and low
background, using general purpose LHC detectors CMS and ATLAS. The
muon trigger of these detectors provides $ \mu$ detection with the
$P_T$ threshold of few $GeV/c$. Though it is optimized for nominal
$pp$ interactions, a similar muon trigger can be successfully used in
heavy ion interactions as well.
As for jet studies, jets with transverse
energies above $50~GeV$ can be recognized with high efficiency
and low background even in central $Pb-Pb$ collisions {\cite{tn12}}.

\vspace{1.0cm}
\section{$Z$ production as a reference process}
\vspace{0.5cm}

As mentioned above, serious normalization problems of the cross sections
measured in heavy ion collisions
can be avoided, if an appropriate reference process is found, and
{\it{ratios}} of the two processes are compared in nucleus-nucleus and
proton-proton interactions. It is most important that the
reference process is not sensitive to final state nuclear interaction
effects, so production of the large invariant mass lepton pair continuum
seems to be an obvious candidate. Explicit calculation in realistic
experimental conditions show, however, that a significant background
coming from heavy quark semileptonic decays and low production rates
of lepton pairs with high invariant mass $M(\mu^+\mu^-) > 100~GeV$
make this process almost useless for the above purpose.

We argue that the production of the $Z$ boson
and its subsequent leptonic decay $Z \rightarrow \mu^+ \mu^-$
should be a good choice for a reference process for a number of reasons:

\begin{itemize}

\item{$Z$ decays weakly into muons, and final state interactions are
clearly negligible in this case;}

\item{the cross section is high enough, and the background can be shown
to be well under control, so that the expected error
can be made sufficiently small;}

\item{the invariant mass of the produced system
$\sqrt{\hat s}=M_Z$ is large enough, not too far from
the average invariant mass
of relevant high $P_T$ jets, $Z+jet$ system and other possible
hard processes of interest,
which makes their production kinematics quite similar;}

\item{momenta of initial partons involved are large enough,
$x_{1,2}\sim \sqrt{\hat s/s} \geq 0.02$, so that the very small $x$ region,
most strongly affected by the initial state nuclear interaction
 like gluon shadowing, is avoided.}

\item{ As the Z-boson is produced mainly by quark-antiquark fusion mechanism,
a very good self-consistency check can be performed, by using quark and
antiquark distribution functions extracted from deep inelastic
lepton-nucleus scattering for describing Z inclusive distributions in
heavy ion collisions.}

\item{ General purpose LHC detectors are capable of
measuring Z production with high efficiency and resolution.}

\end{itemize}

    So, we propose to use the LHC heavy ion facility to
study: i) the ratio of the number of high $P_T$ jets to
the number of detected $Z \rightarrow \mu^+ \mu^-$ decays, and
ii) the ratio of the number of $Z+jet$ events to
the same number of detected $Z$-boson decays,
in various
heavy ion collisions as well as in $pp$ interaction. Note that for the
latter process we expect significant cancellation of initial-state
nuclear effects in the ratio, as initial quarks are involved in both
the numerator and the denominator. Expected statistics
 will allow to measure the variation of these ratios
with the  transverse energy flow and other quantities of interest.

\vspace{1.0cm}
\section{Experimental aspects and results of simulation}
\vspace{0.5cm}

    At LHC heavy ions will be accelerated up to the energies
 $E=E_p \cdot (2Z/A)$ per nucleon pair, where $E_p=7~TeV$ is the
 proton beam energy for LHC.
 In the case of $Pb$ nuclei the energy per nucleon pair will be
 $5.5~TeV$ and the design luminosity for a single experiment is about
 $L \approx 1.0 \times 10^{27}~cm^{-2}s^{-1}$. The event rate for
 $Pb-Pb$ interactions is expected to be in the range of $6-7~ kHz$, but
 only $2-3~\%$ of it will correspond to central collisions.

    The rates of $jet$, $Z + jet$ and $Z$  production
(processes (\ref{jj}-\ref{z})) in nucleus-nucleus
 collisions were obtained from $pp$
 interactions at the same energy using the
 linear A-dependence: $ \sigma_{AA}=A^{2 \alpha} \times
 \sigma_{pp}$, with $\alpha=1.0$.
 The cross-sections in
 $pp$ collisions were evaluated using the PYTHIA 5.7
 Monte-Carlo program {\cite{tn4}}, with
 the default structure function set (CTEQ2L) and the $K$-factor equal to $2$.
 The effects of deflection for quark structure functions with small
 fractional momenta
 in a nucleus relative to a free nucleon, and energy
 losses of high $P_T$ partons in the dense matter were
 not taken into account.

    The charged particle density in mid-rapidity range for the central
 $Pb-Pb$ interactions at LHC is expected to be around $2000-4000$.
 Several Monte-Carlo
 event generators have been created for simulating such collisions.
 HIJING {\cite{tn6}},
 FRITIOF {\cite{tn7}} and  VENUS {\cite{tn8}} are
 often used for nucleus-nucleus collision studies at SPS energies.
 We have used another approach to the simulation of this process,
 implemented in the SHAKER Monte-Carlo program,
 where gross features of soft particle production in heavy ion collisions
 are reproduced using some simple parametrizations.
 The main parameter
 for this generator is the charged particle density in central rapidity
 region, according to which the rates of various types of
 particles are computed.
 Particle production was generated according to
 the flat rapidity distribution in the region $|Y| <2.5$ and
 experimental distributions
 measured at the Tevatron were used as transverse momentum spectra.
 After that, the simulated central heavy-ion collision event was added
 to a hard $pp$ interaction event with
 a high $P_T$ jet and $Z$ production.
 The following kinematical cuts have been applied:
\begin{itemize}
\item {c.m.s. rapidity $Y$ of the $Z$ and the jet should be within the
limits $|Y| < 2.5$;}
\item {transverse momentum of each muon is larger than $20~GeV/c$. Such
muons will be measured with very high efficiency and low background
\cite{tn12};}
\item {transverse energy of the jet is larger than $50~GeV$.
Reconstruction efficiency for jets with
smaller transverse energy is fairly low, and large contamination from
false jets is anticipated; }
\item {transverse momentum of the muon pair is larger than $50~GeV/c$.
This cut eliminates the background from uncorrelated lepton pairs and
maintains the balance of the transverse energy in the event.}
\end{itemize}
 In this kinematical region the contribution of parton
 subprocesses $q+g \to Z+q$ and $q+\bar q \to Z+g$ are $71\%$
 and $29\%$ respectively.
 The Monte-Carlo study has shown that
 for $P_T>50~GeV/c$ the jet recognition efficiency of $95\%$ can
 be achieved, while the
 contribution of false jets can be reduced down to $10\%$ {\cite{kk}}.

    The dominant background of $\mu^+\mu^-$ pairs
in this case comes from two sources: $Z$ production
 with the misrecognition of a fluctuation in the transverse momentum
 flow of soft particles as a jet, and
 QCD processes of heavy flavor pair
 $c \bar c, b \bar b, t \bar t$  production with their subsequent
 semileptonic decays. However, the cut upon the transverse momentum of
 the muon pair
 ${P_T}^{\mu+ \mu-}>50~GeV/c$ significantly reduces both of them,
  so that the
 final contribution of the background is less than $5\%$
 even without using the muon isolation criteria.
 The background due to uncorrelated muon pairs from $\pi$ and $K$
 decays is negligible for ${P_T}^{\mu}>20~GeV/c$.
 The expected number of $jet +Z$ events is around $2000$.

For the same kinematical region the expected number of high $P_T$
 jets in $Pb-Pb$ collisions (process (\ref{jj})) is $ \sim 10^8$,
 while the expected number of detected $Z \rightarrow \mu^+ \mu^-$ decays
 (process (\ref{z})) in the same reaction is around
 $ \sim 5\times 10^4$.
 The luminosity for lighter ions at LHC should be much larger,
 so that numbers of events in case of
 $Ca-Ca$ interactions are expected to be about $100$ times higher for
 the same period of running time, with similar background levels.

\vspace{1.0cm}
\section{Conclusion}
\vspace{0.5cm}

We have shown that the high transverse momentum $Z$-boson production
in association with a hadronic jet  in heavy ion interactions
(reaction (\ref{zj})) is a unique
tool for investigating fast colour charge (mainly light quark)
propagation through dense hadronic matter and searching for the
manifestations of the quark-gluon plasma formation at high energy
densities.
Despite its smaller cross section, which still happens to be
large enough to be
measurable with sufficient accuracy using general purpose LHC detectors,
the proposed reaction can successfully
complement similar studies of high
transverse momentum jet production (reaction (\ref{jj}))
in a number of aspects, and has several important advantages.

In order to compare the rates of various hard processes in different
heavy ion and proton-proton collisions, an appropriate reference
process should be used.
We argue that for a number of theoretical and experimental reasons,
the process of $Z$ boson production and its subsequent muonic decay
(reaction (\ref{z})) is the best choice for such a process satisfying all
necessary conditions.

We would like to thank M.Bedjidian and D.Denegri  for their
interest and helpful discussions.

\end{document}